\documentclass[aps,prl,longbibliography,superscriptaddress,reprint]{revtex4-2}

\usepackage{graphicx}
\usepackage{amsmath}
\usepackage{multirow} 
\usepackage{physics} 
\usepackage{braket}
\usepackage{times}
\usepackage{gensymb}
\usepackage{hyperref} 
\usepackage{amssymb}
\hypersetup{colorlinks=true, citecolor=blue, urlcolor=blue, linkcolor=blue}
\usepackage{soul}
\usepackage{color, xcolor}
\usepackage{textcomp}

\soulregister{\cite}7 % 注册\cite命令
\soulregister{\citep}7 % 注册\citep命令
\soulregister{\citet}7 % 注册\citet命令
\soulregister{\ref}7 % 注册\ref命令
\soulregister{\pageref}7 % 注册\pageref命令

\begin{document}

% figures 9:4 ratio
% different subfigures need to align
% unify format of the various plots

\title{Experimental Proposal on Scalable Radio-Frequency Magnetometer with Trapped Ions}
% Exponential enhancement of estimation precision for multiple parameters with two-mode squeezed state

\author{Yuxiang Huang}
\affiliation{Laboratory of Spin Magnetic Resonance, School of Physical Sciences, Anhui Province Key Laboratory of Scientific Instrument Development and Application, University of Science and Technology of China, Hefei, 230026, China}
\affiliation{Hefei National Laboratory, University of Science and Technology of China, Hefei 230088, China}

\author{Wei Wu}
\affiliation{Laboratory of Spin Magnetic Resonance, School of Physical Sciences, Anhui Province Key Laboratory of Scientific Instrument Development and Application, University of Science and Technology of China, Hefei, 230026, China}

\author{Qingyuan Mei}
\affiliation{Laboratory of Spin Magnetic Resonance, School of Physical Sciences, Anhui Province Key Laboratory of Scientific Instrument Development and Application, University of Science and Technology of China, Hefei, 230026, China}

\author{Yiheng Lin} \email{yiheng@ustc.edu.cn}
\affiliation{Laboratory of Spin Magnetic Resonance, School of Physical Sciences, Anhui Province Key Laboratory of Scientific Instrument Development and Application, University of Science and Technology of China, Hefei, 230026, China}
\affiliation{Hefei National Research Center for Physical Sciences at the Microscale, Hefei 230026, China}
\affiliation{Hefei National Laboratory, University of Science and Technology of China, Hefei 230088, China}

%\author{Jiangfeng Du} \email{djf@ustc.edu.cn}
%\affiliation{CAS Key Laboratory of Microscale Magnetic Resonance and School of Physical Sciences, University of Science and Technology of China, Hefei 230026, China}
%\affiliation{CAS Center for Excellence in Quantum Information and Quantum Physics, University of Science and Technology of China, Hefei 230026, China}
%\affiliation{Hefei National Laboratory, University of Science and Technology of China, Hefei 230088, China}
%\affiliation{Institute of Quantum Sensing and School of Physics, Zhejiang University, Hangzhou 310027, China}

\begin{abstract}

Quantum magnetometry represents a fundamental component of quantum metrology, where trapped-ion systems have achieved pT/$\sqrt{\text{Hz}}$ sensitivity in single-ion radio-frequency magnetic field measurements via dressed states based dynamical decoupling. Here we propose a scalable trapped-ion magnetometer utilizing the mixed dynamical decoupling method, combining dressed states with periodic sequences to suppress decoherence and spatial magnetic field inhomogeneity. With numerical simulations for a $10^4$ ion system with realistic experimental parameters, we demonstrate that a sensitivity of 13 fT/$\sqrt{\text{Hz}}$ for the radio-frequency field could be reached. Such a sensitivity could be obtained via robust resilience to magnetic field drift noise and inhomogeneity, where coherence time could be extended to the order of several minutes on average. This method enables scalable trapped-ion magnetometry, demonstrating its potential as a robust and practical solution for advancing quantum sensing applications.

\end{abstract}

\maketitle

Quantum metrology~\cite{Quantum_Metrology} crucially drives modern technological advancement, with quantum magnetometers serving as core tools for ultra-sensitive magnetic field detection. These devices are vital across diverse applications, ranging from imaging~\cite{Imaging1,Imaging2}, materials characterization~\cite{Material_Characterization} and probing fundamental physics beyond the Standard Model~\cite{Exotic_Interaction,Axion_Search}. 
Developed platforms include atomic magnetometer working in the spin exchange relaxation-free (SERF) regime ~\cite{SERF1,SERF2,SERF3,SERF4, SERF_subfT}, superconducting quantum interference device (SQUID) ~\cite{Squid1,Squid2,Squid3,Squid4}, nitrogen-vacancy (NV) center ~\cite{Imaging2,NV2,NV3,NV4} and trapped-ion magnetometer~\cite{Ion_Lockin,Ion_Cai,Ion_DCentangle,Ion_Cui}. Current sensitivities reach sub-fT/$\sqrt{\text{Hz}}$ ~\cite{Squid1,Atomic_subfT,SERF_subfT}, enabling measurements across a broad frequency spectrum ranging from DC~\cite{Ion_DCentangle,SERF_subfT} to GHz~\cite{Imaging2, Atom_GHz} via complementary sensing mechanisms. 

Among these platforms, single trapped-ion magnetometer achieves pT/\(\sqrt{\text{Hz}}\) \cite{Ion_Cai} sensitivity with more than two seconds coherence time in the radio-frequency (RF) band. To further improve the sensitivity of magnetometry, it is necessary to further extend the coherence time and increase the number of ions. On one hand, trapped-ion qubits exhibit coherence times reaching from tens of minutes~\cite{Ion_10min} to over one hour~\cite{Ion_1h} by leveraging advanced dynamical decoupling to suppress environmental noise. However, integration of these dynamical decoupling protocols onto trapped-ion magnetometers remains to be developed and demonstrated. On the other hand, the trapped-ion system also exhibits strong scalability, where three-dimensional Coulomb crystals containing approximately $10^4$ or even $10^5$ ions ~\cite{Ion_10^4,Ion_10^5} can be realized in Paul traps via sympathetic cooling. Such large ion ensembles have been proposed and successfully applied to quantum precision metrologies such as atomic clocks~\cite{Ionclock1,Ionclock2,Ionclock3,Ionclock4}, paving the pathway for scalable trapped-ion magnetometry. However, the large ion crystal is more sensitive to spatial magnetic field inhomogeneity, introducing inconsistent transition frequencies across ions, as recently revealed in scalable ion clocks~\cite{CdClock2015,CdClock2021}. Similarly, this issue compromises scalability of trapped-ion magnetometry, since non-uniform frequency detuning systematically makes signals deviate from ideal response characteristics.

In this work, we utilize a mixed dynamic decoupling (MDD)~\cite{mixedDD} scheme, a hybrid class of dynamic decoupling protocols~\cite{Sensing_RMP, DD, DDexp_solid}, to enhance sensitivity performance in RF magnetic field sensing with trapped-ion systems. Our protocol stems from the integration of pulsed dynamic decoupling ~\cite{PDD1} with the dressed-state method~\cite{Ion_Cai,Ion_Cui,Ion_dressed} or continuous dynamic decoupling  ~\cite{CDD1}. Mechanistically, MDD achieves noise suppression in both temporal and spatial domains. Temporally, such a method prolongs the coherence time of atomic sensors, while spatially enhancing inherent robustness against magnetic field inhomogeneities on large trapped-ion crystal. These two features enable reliable operation in scalable systems. Numerical simulations demonstrate that even under an inhomogeneous field in realistic experimental conditions, the proposed MDD protocol extends coherence times by two orders of magnitude, achieving RF magnetic field sensitivities to 13 fT/\(\sqrt{\text{Hz}}\) at more than 1 minute coherence time for RF magnetometry with $10^4$ ions.

\begin{figure}[t] 
    \centering
    \includegraphics[width=1.0\columnwidth]{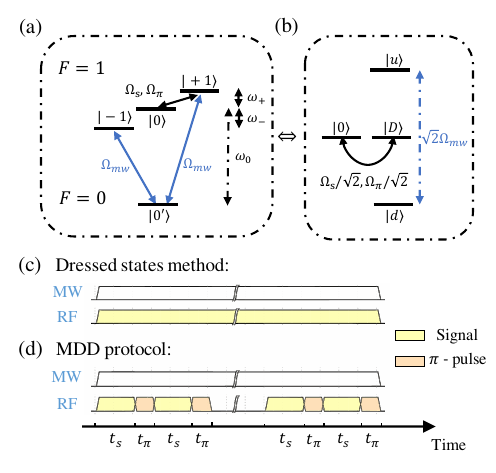}
    \caption{\textbf{$^{171}$Yb$^+$ level scheme and time sequences for dressed states and MDD methods in RF magnetometry.} (a) Hyperfine levels of the $^{171}$Yb$^+$ ground states used as the atomic sensor. Two resonant MW fields with identical Rabi frequency $\Omega_{mw}$ are applied, with the signal RF field (Rabi frequency $\Omega_s$) and echo RF field (Rabi frequency $\Omega_\pi$) driving the transition from $\left| 0 \right\rangle$ to $\left| +1 \right\rangle$. (b) The corresponding RF transition is equal to the transition from $\left| 0 \right\rangle$ to the dressed state $\left| D \right\rangle$ with Rabi frequency $\Omega_s/\sqrt{2}$ (or $\Omega_\pi/\sqrt{2}$) in the dressed states basis \{$\left| 0 \right\rangle$, $\left| D \right\rangle$, $\left| u \right\rangle$, $\left| d \right\rangle$\}. (c) For the dressed states method, both resonant MW fields are continuously applied during the period when the signal RF field drives the transitions. (d) In the MDD method, a CPMG-concatenated sequence is implemented. Specifically, several RF $\pi$-pulses are introduced during the RF signal intervals to induce a population inversion between the $\left| 0 \right\rangle$ and $\left| D \right\rangle$ states. Both (c) and (d) omit the state preparation at the beginning and the detection process at the end of the sequence.}
    \label{fig:sequences}
\end{figure}

{\it Experimental scheme.} We consider an atomic sensor utilizing four hyperfine levels in the $^2S_{1/2}$ manifold of $^{171}$Yb$^+$ ion, as illustrated in Fig. \ref{fig:sequences}(a). These states are defined with $\ket{0'}\equiv\ket{F=0, m_F=0}$, $\ket{0}\equiv\ket{F=1, m_F=0}$, $\ket{-1}\equiv\ket{F=1, m_F=-1}$ and $\ket{+1}\equiv\ket{F=1, m_F=+1}$. The hyperfine splitting between $\ket{0'}$ and $\ket{0}$ is denoted by $\omega_0$. An external static magnetic field $B_z$ applied along the quantization axis lifts the degeneracy of the $F=1$ states, splitting them into three $m_F$ sub-levels with frequency separations $\omega_+$ and $\omega_-$.

In environments with low-frequency magnetic field noise (e.g., from temperature or current drifts), accurately measuring the amplitude of RF magnetic field signals resonant with $\omega_+$ or $\omega_-$ requires robust coherence protection. Previous single-ion magnetometry studies~\cite{Ion_Cai, Ion_Cui} employed a dressed states method, with two microwave (MW) fields resonant with $\ket{0'}\leftrightarrow\ket{-1}$ and $\ket{0'}\leftrightarrow\ket{+1}$ transitions, both having the same Rabi frequency $\Omega_{mw}$. This setup produces a dressed state $\ket{D} = (\ket{+1}-\ket{-1})/\sqrt{2}$, designed to be insensitive to first-order magnetic field fluctuations in the transition of $\ket{0}\leftrightarrow\ket{D}$. Building upon such a foundation, we suppress the effects from low-frequency noise and spatial field inhomogeneity in scalable trapped-ion systems by integrating a Carr-Purcell-Meiboom-Gill (CPMG) sequence~\cite{CPMG} with the dressed states method. This hybrid approach suppresses residual low-frequency magnetic fluctuations and mitigates the dephasing effects of inhomogeneous magnetic fields, both of which have been critical limitations for scalable trapped-ion magnetometry.
When the RF signal field to be measured is applied, we tune the static magnetic field $B_z$ so that the $\ket{0}\leftrightarrow\ket{-1}$ (or $\ket{0}\leftrightarrow\ket{+1}$) transition is near resonant with the signal RF frequency. This drives Rabi oscillation between $\ket{0}$ and $\ket{D}$ with effective Rabi frequencies $\Omega_0=\Omega_s/\sqrt{2}$ and $\Omega_0'=\Omega_{\pi}/\sqrt{2}$, where $\Omega_s$, $\Omega_\pi$ denote the Rabi frequencies of the measured RF signal and CPMG $\pi$-pulses, respectively. The Hamiltonian of the system is described as $H_{sys}$, where 
\begin{equation}
\begin{aligned}
    H_{sys} &= H_0 + H_{mw} + H_{rf}
     \\    
    H_0 &= -\omega_0 \ket{0'}\bra{0'} + \omega_+  \ket{+1}\bra{+1} -\omega_- \ket{-1}\bra{-1}\\
    H_{mw} &= \Omega_{mw}\left(\ket{-1}\bra{0'}e^{i (\omega_0-\omega_-)t} \right.\\
    &\qquad\qquad \left.+ \ket{+1}\bra{0'}e^{i (\omega_0+\omega_+)t} + H.c. \right) \\
    H_{rf} &= \left[ \Omega_{s} f_1(t) cos(\omega_+ t + \phi_s) \right.  \\
    &\quad \left.+ \Omega_\pi f_2(t) cos(\omega_+ t + \phi_\pi)  \right]    \times\left( \ket{+1}\bra{0}+ H.c. \right)
\end{aligned}
\end{equation}
Here, $H_0$, $H_{mw}$, $H_{rf}$ represent the atom levels Hamiltonian, microwave dressed field Hamiltonian and RF driving Hamiltonian, respectively. The phases of RF signal and RF $\pi$-pulse are set as $\phi_s$ and $\phi_\pi$.  And $f_1(t)$, $f_2(t)$ are the modulation functions of RF signal and $\pi$-pulse related to CPMG sequences. Specifically, $f_{1,(2)}(t)=1$ and $f_{2,(1)}(t)=0$ when the RF signal (or $\pi$-pulse) operates.

After applying the rotating wave approximation and projecting into the dressed states basis, the interaction Hamiltonian is given by
\begin{equation}
\begin{aligned}
    H_{int} &= \frac{\Omega_{mw}}{\sqrt{2}}\left(\ket{u}\bra{u} -  \ket{d}\bra{d}\right) \\
    &\qquad+ \frac{\Omega_{s}}{\sqrt{2}}f_1(t)\left( \ket{D}\bra{0} e^{i\phi_s}+ H.c. \right) \\
    &\qquad+ \frac{\Omega_{\pi}}{\sqrt{2}}f_2(t)\left( \ket{D}\bra{0} e^{i\phi_\pi}+ H.c. \right)
\end{aligned}
\end{equation} 
 which leads to the interaction picture basis of dressed states, as shown in Fig. \ref{fig:sequences}(b) with states $\ket{u}=(\ket{B}+\ket{0'})/\sqrt{2}$,  $\ket{d}=(\ket{B}-\ket{0'})/\sqrt{2}$ defined with $\ket{B}=(\ket{+1}+\ket{-1})/\sqrt{2}$. The states $\ket{u}$ and $\ket{d}$ are separated from $\ket{D}$ by an energy of $\pm \Omega_{mw}/\sqrt{2}$ respectively.
 The MDD method consists of the following sequences: (evolution - $\pi$-pulse - evolution - $\pi$-pulse)$^{\bigotimes N}$, as illustrated in Fig. \ref{fig:sequences}(d). The evolution part is executed in the form similar to the dressed states method, with RF signal and dressed microwave applied over a time interval $t_{s}$ presented in Fig. \ref{fig:sequences}(c). The $\pi$-pulses generated by an additional RF field are applied with a stronger Rabi frequency $\Omega_\pi\gg\Omega_s$ and the same frequency as the RF signal. These $\pi$-pulses lead to the exchange of population between $\ket{0}$ and $\ket{D}$ states, forming CPMG sequences. We note that the RF signal can only be fully preserved when its phase aligns with that of the $\pi$-pulse, such that $\phi_s=\phi_\pi$. Conversely, components with phase discrepancy from the $\pi$-pulse are regarded as noise and filtered out by the CPMG sequences. This characteristic therefore allows for the estimation of the phase of signal in addition to its amplitude by performing the MDD method.

 \begin{figure}[htbp]
    \centering
    \includegraphics[width=\columnwidth]{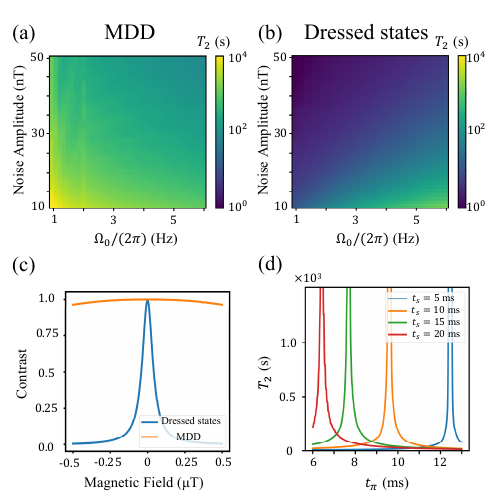}
    \caption{\textbf{Coherence time $T_2$ of single atomic sensor under dressed states and MDD methods.} (a) Coherence time $T_2$ of a single atomic sensor under varying magnetic field noise amplitudes $\Delta B$ and RF signal strengths $\Omega_s$ for the MDD method, showing an obvious enhancement compared to the dressed states method. The Rabi frequency of the $\pi$-pulse $\Omega_\pi$ is set to satisfy $t_\pi=6.3$ ms and the effective evolution time is set to $t_s=20$ ms here to achieve the optimal coherence time. (b) Coherence time $T_2$ for the dressed states method, where the Rabi frequencies of two dressed microwave fields are maintained at $\Omega_{mw}=2\pi\times25$ kHz, consistent with that in (a). (c) For the single ion, MDD method demonstrates a better robustness compared to dressed states method under magnetic field drifts. (d) Optimal coherence time $T_2$  for the MDD method as a function of the $\pi$-pulse duration $t_\pi$ (or equivalently, the Rabi frequency $\Omega_\pi$) under fixed effective signal strength $\Omega_0=2\pi\times$1 Hz and magnetic field noise magnitude $\Delta B=0.05$~\textmu T. The coherence time $T_2$ reaches an optimal value for specific durations of the $\pi$-pulse, illustrating the impact of the duty cycle of the sequence on coherence enhancement.}
    \label{fig:single_ion_T2}
\end{figure} 
\par

{\it Numerical simulation for a single-ion magnetometer.} We performed numerical simulations to assess the effectiveness of the MDD method in enhancing the coherence time $T_2$ of a single atomic sensor. Since the sensor sensitivity scales with $T_2^{-1/2}$, the enhancement of $T_2$ directly improves sensitivity in RF magnetic field measurements.
Attributing shot-to-shot magnetic field fluctuations as the sole noise source from temperature or current drifts, we sampled the magnetic field amplitude of each trial from a Gaussian distribution centered at $B_0 = 0.765$ mT with variable noise amplitude $\Delta B$ (FWHM). Figure \ref{fig:single_ion_T2} compares $T_2$ for dressed states and MDD methods across different parameters, where $T_2$ is defined as the time when signal contrast of the Rabi oscillation decays to $1/e$.

Setting the Rabi frequencies of $\pi$-pulses $t_\pi=6.3$ ms and the single effective evolution time $t_s=20$ ms, we find that MDD method extends the coherence time by nearly three orders of magnitude compared to the existing dressed states method under the same RF signal strengths $\Omega_s$ and magnetic field fluctuations, as shown in Fig. \ref{fig:single_ion_T2}(a) and (b).  In Fig. \ref{fig:single_ion_T2}(b), we show the coherence time of the dressed states method in the same parameters ranges within several seconds, consistent with previous works ~\cite{Ion_Cai,Ion_DCentangle,Ion_Cui}. Note that smaller noise intensities and larger signal strengths lead to prolonged contrast for this method. In contrast, Fig. \ref{fig:single_ion_T2}(a) reveals that the Rabi oscillation for the MDD method has at least hundred-fold improvement than that of the dressed states method for a broad parameter setting. The improvement is even more significant at low signal strengths. 
Such a capability of mixed dynamical decoupling is due to enhanced mitigation of the second-order magnetic field noises between states $\ket{0}$ and $\ket{D}$. In Fig.~\ref{fig:single_ion_T2}(c), we compare single-ion magnetometry performance between dressed states and MDD methods under fixed magnetic field drift. The MDD method outperforms the dressed states method and maintains higher contrast, enabling enhanced robustness against temporal magnetic field noise and spatial inhomogeneity. For optimal coherence time, the measurement duty cycle is set to align with MDD’s optimal case as described before. 
Notably, the MDD method exhibits optimal echo strengths $t_\pi$ at different evolution time $t_s$ and we plot this relationship in Fig. \ref{fig:single_ion_T2}(d). This phenomenon is absent in the two-level system, which arises from competition between the $\pi$-pulse intensity $\Omega_\pi$ and the microwave intensity $\Omega_{mw}$ used to generate the dressed states. Excessive $\Omega_\pi$ excites the population leakage from the subspace $\{\ket{0}, \ket{D} \}$ which is not sensitive to the first-order magnetic field noise, into the noise sensitive sub-spaces $\{\ket{u}, \ket{d} \}$. This leakage influences the MDD method performance and presents limits on further improving the coherence time $T_2$.

\begin{figure*}[htbp]
    \centering
    \includegraphics[width=2\columnwidth]{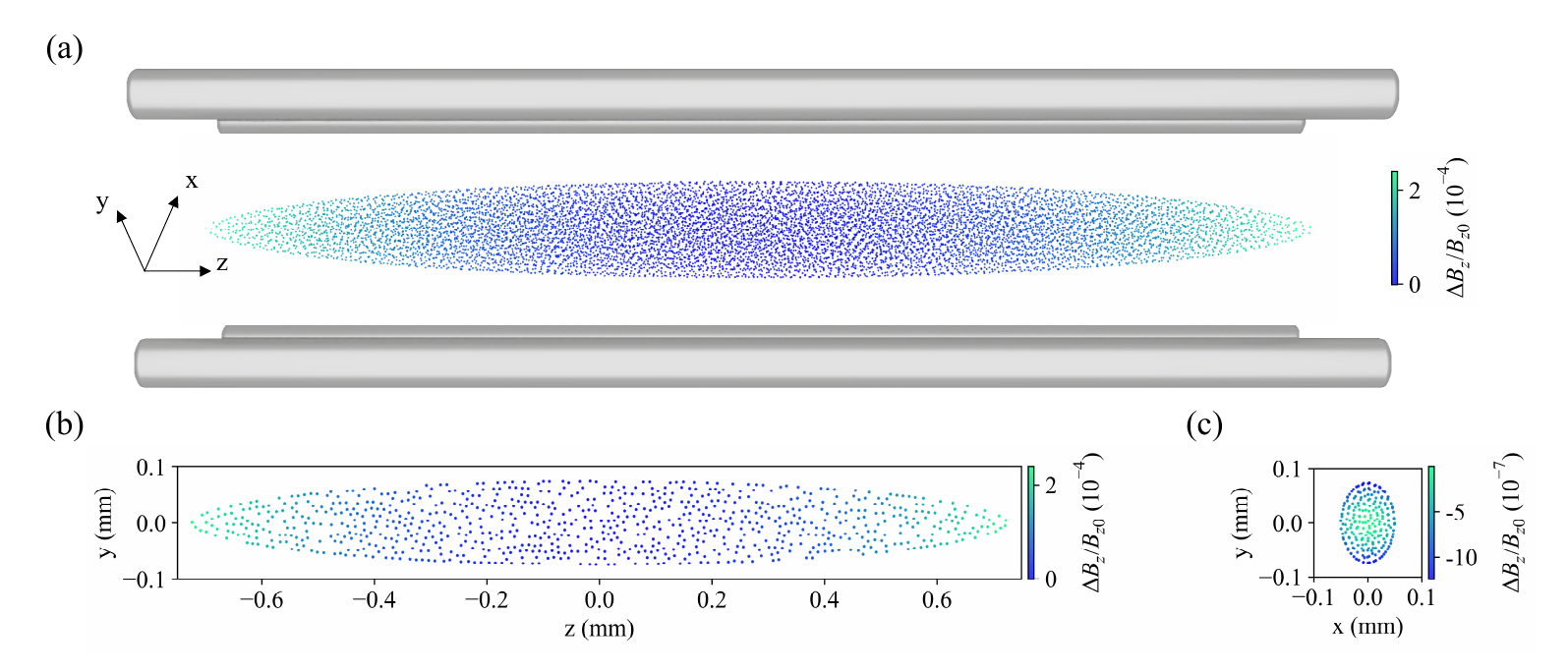}
    \caption{\textbf{Axial magnetic field inhomogeneity of a Coulomb crystal with 10,000 ions in a Paul trap.} (a) A Coulomb crystal consisting of 10,000 ions trapped in a Paul trap, sized 1.45 mm $\times$ 0.15 mm $\times$ 0.10 mm, with assumed trap frequencies of $\{ \omega_x, \omega_y, \omega_z \}=2\pi\times\{$0.7, 0.58, 0.12$\}$ MHz. The color describes the inhomogeneous magnetic field experienced by each ion. (b) Axial magnetic field inhomogeneity near the x=0 plane, and the crystal experiences magnetic field non-uniformity of about $10^{-4}$ along the axial direction. (c) Axial magnetic field inhomogeneity near the z=0 plane, corresponding to a radial inhomogeneity of $10^{-6}$. }
    \label{fig:crystal}
\end{figure*}

{\it Towards multi-ion magnetometer.} Having demonstrated that the MDD method significantly enhances the coherence time $T_2$ in single-ion configurations, we now explore the feasibility of scaling up atomic sensors for enhanced sensitivity. A key approach to boosting trapped-ion magnetometer sensitivity is increasing the number of atomic sensors $N_i$, as sensitivity improves according to the relationship $S\propto 1/\sqrt{N_i}$. This scaling arises from the standard quantum limit, where $N_i$ uncorrelated sensors collectively enhance the measurement signal-to-noise ratio by a factor of $\sqrt{N_i}$, corresponding to a reduction in the population uncertainty of the ensemble \cite{ProjectionNoise}. %It is important to note that this improvement relies on the ideal assumption that the signals from all ions add coherently without introducing additional correlations in the projection noise.
Against this background, we introduce a scalable trapped-ion system to numerically investigate how spatial magnetic field inhomogeneity impacts the performance of a multi-ion magnetometer.

\begin{figure}[htbp]
    \centering
    \includegraphics[width=\columnwidth]{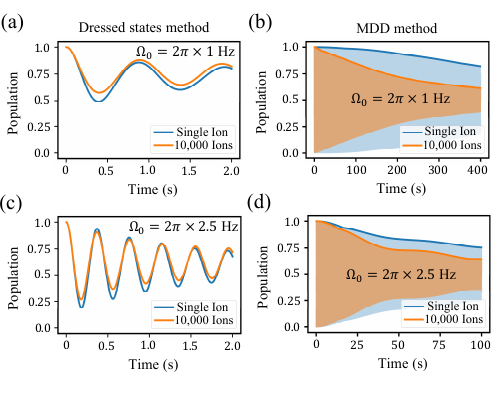}
    \caption{\textbf{Effect of magnetic field inhomogeneity with shot-to-shot temporal noise on measurement.} (a) Rabi oscillations under magnetic field inhomogeneity with temporal noise using the dressed states method. The coherence time of the single ion at the center of the crystal (blue curve) is $T_2=0.62$ s, while the coherence time of average 10,000 ions (orange curve) is $T_2=0.36$ s with effective signal strength  $\Omega_0=2\pi\times 1$ Hz. (b) Similar to (a), but for the MDD method. The contrast of the Rabi oscillations remains perfect, and the coherence time of the single ion increases to  $T_2=620$~s, with the average coherence time of $10,000$ ions reaching  $T_2=250$~s. (c)(d) Same as (a)(b), except with different effective signal strength $\Omega_0=2\pi\times 2.5$ Hz. The coherence times in different cases become $T_2=1.4$~s (single ion) and $T_2=1.2$~s (10,000 ions) when using dressed states method, and to be $T_2=150$ s (single ion) and $T_2=75$ s (10,000 ions) when using MDD method. Here we apply $\Omega_{mw}=2\pi\times 25$ kHz, $t_s=20$ ms and $t_\pi=6.3$ ms.}
    \label{fig:multi_T2}
\end{figure} 

Specifically, we simulate the structure of a $10^4$-ion Coulomb crystal confined in a traditional linear Paul trap using molecular dynamics calculations similar to previous work \cite{LargeCrystal, Crystalwyk}. The trap frequencies are set as $\{ \omega_x, \omega_y, \omega_z \}=2\pi\times\{0.7, 0.58, 0.12\}$ MHz. Under the pseudo-potential assumption, the simulated crystal exhibits dimensions of $1.45$ mm $\times$ $0.15$ mm $\times$ $0.10$ mm, as shown in Fig. \ref{fig:crystal}(a).
Then we characterize spatial magnetic field inhomogeneity within the Coulomb crystal. To simulate as closely as possible to the experimental parameters, we consider a pair of Helmholtz coils (radius = 5 cm, spacing = 10 cm) to generate a static magnetic field aligned with the ion crystal’s axial direction. The axial magnetic field at the coil center is defined as $B_{z0}$, against which the amplitude $B_{z}$ at other positions is normalized. The magnetic field inhomogeneity is quantified as $(B_z-B_{z0})/B_{z0}=\Delta B_z / B_{z0}$. As shown in Fig. \ref{fig:crystal}(b) and (c), the axial inhomogeneity reaches $10^{-4}$, while radial inhomogeneity is an order of magnitude lower at $10^{-6}$.  Additionally, we assume each ion experiences shot-to-shot magnetic field noise from environments, which is sampled from a Gaussian distribution with noise amplitude $\Delta B=0.05$ \textmu T around an average field of $B_0=0.765$ mT.
Based on this characterization, we evaluate the MDD method’s resilience to such inhomogeneities.

We simulate the impact of static magnetic field inhomogeneity with temporal noise on multi-ion magnetometers using both dressed states and MDD methods, as shown in Fig.  \ref{fig:multi_T2}.  In both panels, blue curves represent ions locating at the center of the crystal, where magnetic field inhomogeneity is set to zero as a reference.  Orange curves show the averaged signal from all 10,000 ions, capturing collective responses to both spatial inhomogeneity and shot-to-shot temporal noise. The vertical axis tracks the population of initial state $\ket{0}$ during evolution. For the dressed states method, the contrast of Rabi oscillation decays over the duration of coherence time of less than 2 s, for instance with  $\Omega_0/2\pi$ of 1 and 2.5~Hz respectively, as shown in Fig. \ref{fig:multi_T2}(a) and (c). In contrast,  the MDD method maintains high contrast oscillations, achieving $T_2>60$~s for the same drive strengths $\Omega_0$ as above, as depicted in Fig. \ref{fig:multi_T2}(b) and (d). This result demonstrates MDD's robustness against spatial variations and temporal noise across different signal strengths, highlighting its potential for noise suppression and precision enhancement in scalable trapped-ion magnetometers with a wide dynamical range. 

Scaling up from single-ion to multi-ion magnetometry introduces two residual noises: spatial variations in RF coil amplitudes and dressed microwave Rabi frequencies. The first noise type, comparable to field inhomogeneity on the order of $10^{-4}$, is included in our simulation. The second can be technically suppressed below 0.1\% for millimeter-scale crystals and has a negligible effect on $T_2$ during measurement process. We further evaluated the effect of micromotion on Rabi frequencies under a first-order approximation, where the Rabi frequency $\Omega$ is modulated as $\Omega J_0(kx_m)$ \cite{MicroGates, MicroChen}, with $J_0$ denoting the zeroth-order Bessel function, $k$ the wavevector of the driving field, and $x_m$ the micromotion amplitude. Numerical simulations indicate a maximum radial micromotion amplitude of approximately 30~µm in our configuration. Assuming radially positioned microwave and RF coils, this amplitude results in a Rabi frequency deviation below 0.01\% for the 12.6~GHz microwave transition and below $10^{-8}$ for RF transitions. Thus, micromotion-induced fluctuations in Rabi rates are negligible under the proposed experimental conditions.
The stability of large Coulomb crystals is ensured by a long linear quadrupole trap setup, which provides a sufficiently large and stable confinement region. The feasibility of confining and coherently controlling crystals of up to $10^4$ ions has been demonstrated in previous trapping and Ramsey experiments \cite{CdClock2015, CdClock2021}. Residual RF heating can be suppressed via active cooling techniques \cite{MDLinear, MDPlanar}, and collisions with background gas are mitigated using a cryogenic environment, thereby preserving crystal stability and coherence over extended measurement durations.

\begin{figure}[h]
    \centering
    \includegraphics[width=\columnwidth]{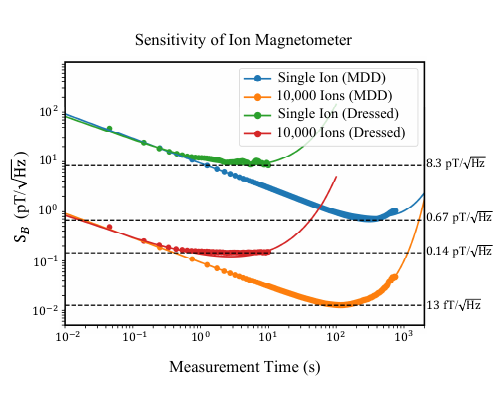}
    \caption{\textbf{Sensitivity of RF magnetometer with different methods.} Enhancements to the standard quantum limit sensitivity are demonstrated through measurement time optimization and scalable atomic sensor configurations. The scatters are plotted according to simulation from sampling, and the solid curves represent the fitting result assuming the decoherence can be approximated by an exponential power law $e^{-(t/T_2)^n}$. As a result, an RF magnetometer using dressed states method with a single ion achieves an optimal sensitivity of 8.3 pT/$\sqrt{\text{Hz}}$ , while the MDD method yields 0.67 pT/$\sqrt{\text{Hz}}$ for a single sensor configuration. In a scalable system with 10,000 ions, both two methods exhibit improved sensitivity, although the coherence times are slightly reduced. Specifically, Dressed states method reaches 0.14 pT/$\sqrt{\text{Hz}}$, and the MDD method achieves 13 fT/$\sqrt{\text{Hz}}$ under spatial inhomogeneity and temporal noise. }
    \label{fig:sensitivity}
\end{figure}

To estimate the achievable magnetic field measurement sensitivity based on the simulation, we first derive the shot-noise-limited sensitivity for the measurement of effective Rabi frequency $\Omega_0$ between the dressed state $\ket{D}$ and state $\ket{0}$ is given by \cite{Ion_Cai}
\begin{equation}
S = \frac{\Delta P}{\left| \frac{\partial P(t) / \partial t}{\Omega_0} t \right| \sqrt{N_i}} \sqrt{T_{tot}}
\end{equation}
where $\Delta P=\sqrt{P(1-P)/n}$ is the standard deviation of the state $\ket{0}$ population $P$ over $n$ repetitions,  $t$ denotes the evolution time and $N_i$ is the number of ions. The factor $\partial P(t) / \partial t$ represents the slope of Rabi oscillations. $T_{tot}=n[(1+t_\pi/t_s)t+t_{add}]$ is the total time needed for $n$ experimental measurements with additional time cost $t_{add}=20$ ms including cooling, detection and state preparation. 
The sensitivity of RF magnetic field amplitude can be calculated through $S_B = \frac{\hbar}{\sqrt{2}\mu_B} S $.
Fig. \ref{fig:sensitivity} illustrates how multi-ion magnetometers enhance sensitivity by increasing $N_i$ and extending the coherence time. The scatter points in the figure are calculated from the simulation data at the maximum Rabi oscillation slopes, while the solid curves are generated by fitting the Rabi oscillation contrast by an exponential power law $e^{-(t/T_2)^n}$ and then using the fitting parameters to calculate the corresponding sensitivity lines. The parameters used for simulating the MDD method are identical to those in Fig. \ref{fig:multi_T2}(b), while the parameters employed for simulating the dressed states are similar to those in Fig. \ref{fig:multi_T2}(a), except that the signal strength is set as $\Omega_0=2\pi\times 5$ Hz here. By extending $T_2$ to the minute scale, the MDD method enables a projected sensitivity of 13 fT/$\sqrt{\text{Hz}}$ for RF magnetic field sensing with $10^4$ ions. The slight discrepancy between the scatter points and the solid curves likely stems from the randomness of shot-to-shot noise in the simulations, as well as the imperfect match between the exponential fitting function and the the Rabi oscillation under shot-to-shot noise.

In conclusion, we propose and analyze a mixed dynamical decoupling method for RF magnetic field sensing, demonstrating its ability to extend coherence time in single- and multi-ion systems with enhanced robustness against spatial and temporal magnetic noises. 
This work establishes a robust foundation for advanced magnetometry techniques, showcasing how the MDD method can enhance sensitivity and be adapted to a scalable sensor architecture across diverse frequency bands.
Furthermore, the spatial extent of the large ion crystal further offers a natural capability for probing magnetic field gradients. This can be realized by partitioning the crystal into independently addressed sub-ensembles using additional electrodes, or by resolving and comparing signals from distinct spatial regions, which enables direct extraction of magnetic field gradients across the crystal.
Our protocol may be beneficial to a range of applications, including searches for exotic interaction between spins and dark matter fields, and the enhancement of coherence and RF field sensitivity in other scalable quantum systems, such as rare-earth ions in solids, and neutral atom arrays trapped via optical tweezers.

\begin{acknowledgments}
\emph{Acknowledgments.---}
We thank J. Cai, H. Qin, X. Ma for helpful discussion. This work was funded by the
 National Natural Science Foundation of China (Grant No.~92165206) and Innovation Program for Quantum Science and Technology (Grant No.~2021ZD0301603).

\end{acknowledgments}

{\it Data availability.} The data supporting findings in this manuscript will be available from the contact author upon reasonable request.

\end{document}